# Electron-Doping Induced Semiconductor to Metal Transitions in ZrSe$_2$ Layers via Copper Atomic Intercalation

*Zahir Muhammad[1], Kejun Mu[1], Haifeng Lv[2], Chuanqiang Wu[1], Zia ur Rehman[1], Muhammad Habib[1], Zhe Sun[1,*], Xiaojun Wu[2] and Li Song[1,*]*

[1]National Synchrotron Radiation Laboratory, CAS Center for Excellence in Nanoscience, CAS Key Laboratory of Strongly-coupled Quantum Matter Physics, University of Science and Technology of China, Hefei, Anhui, 230029, China.

[2]CAS Key Laboratory of Materials for Energy Conservation, Synergetic Innovation Centre of Quantum Information & Quantum Physics, CAS Center for Excellence in Nanoscience, and Department of Material Science and Engineering, University of Science and Technology of China, Hefei, Anhui 230026, China.

**ABSTRACT:** Atomic intercalation in two dimensional (2D) layered materials can engineer the electronic structure at the atomic scale, bringing out tunable physical and chemical properties which are quite distinct in comparison with pristine one. Among them, electron-doped engineering induced by intercalation is an efficient route to modulate electronic states in 2D layers. Herein, we demonstrate a semiconducting to the metallic phase transition in zirconium diselenide ($ZrSe_2$) single crystal via controllable incorporation of copper (Cu) atoms. Combined with first-principles density functional theory (DFT) calculations, our angle resolved photoemission spectroscopy (ARPES) characterizations clearly revealed the emergence of conduction band dispersion at M/L point of Brillouin zone due to Cu-induced electron doping in $ZrSe_2$ interlayers. Moreover, the field-effect transistor (FET) fabricated on $ZrSe_2$ displayed a n-type semiconducting transport behavior, while the Cu-intercalated $ZrSe_2$ posed linear $I_{ds}$ vs $V_{ds}$ curves with metallic character shows n-type doping. The atomic intercalation approach has high potential for realizing transparent electron-doping systems for many specific 2D-based electronics.

Layered transition metal dichalcogenides (TMDCs) $MX_2$ (M = Mo, W, Zr, Hf; X = S, Se) semiconductors have captivated widespread interest due to their significant superior physical and electrical properties.[1-6] The crystalline forms of 2D materials have been the subject of intense investigation owing to their unique physical properties[7], such as direct electronic bandgap in the single-layer limit[8-12], valence band (VB) splitting[12-16], valley degree of freedom[17-21], and excitonic nature of optical band spectra[22-26]. Artem *et al.* recently observed the, the co-occurrence of exchange splitting and giant SOC in graphene at the same time with good magnetic characteristics[27]. Thus, TMDCs are considered as an ideal materials for p–n junctions[28], field–effect transistors with high mobility of electrons[29-31], memory and switching devices[32], optoelectronics and photovoltaic applications[33]. These layered materials can also be found in different forms of hybrid or in heterostructure forms.[34-37] Particularly, the atomic phase engineering can result in observation of tremendous novel physical and electronic phenomena.[38-42] Among them, the intercalation of electron-donating atoms has been an effective method to modulate intrinsic electronic properties, such as shifting the fermi level to higher or lower energy[43], changing band gap characteristics[44-46] and inducing phase transitions.[47,48] Morosan *et al.* reported that the phase transition arising at Fermi level in Cu-intercalated $TiSe_2$ could give rise to the new superconducting state in $TiSe_2$.[49] Topological phase transition occurred in $BiTl(S_{1-\delta}Se_\delta)_2$ and $Bi_2Se_3$, as well as superconductivity in $Cu_xZrTe_{2-y}$ with different doping ratio of Se, Co-phthalocyanine and Cu atoms respectively.[50-52] Similarly, Tsipas *et al.* recently observed the semimetallic phase with massless Dirac Fermions is in $ZrTe_2$.[53] The structure and electronic properties of the materials can be ascribes as a semiconductor or a semimetallic states with a small indirect or direct gap can be suppressed by intercalation or elemental doping.[3, 16, 18-22, 31, 49] Likewise, the semiconductor to metal transition was witnessed $MoTe_2$ by strain or structural and electronic phase transition by W substitution.[54] In principle, the states of quantum matter: charge density waves, insulators, and superconductors can be ascribed from the spectral energy gap with different bandgap structures.[55-57] Therefore, the fundamental changes in the band structure can provide vital information for considering and controlling physical and electronic properties.[58]

In this particular work, we synthesized Cu-intercalated zirconium diselenide ($ZrSe_2$) single crystals via chemical vapor transport technique (CVT), and investigated the change in electronic properties in comparison with pristine $ZrSe_2$ by combining synchrotron-based Angle Resolved Photoemission Spectroscopy (ARPES) with first-principles density functional theory (DFT) calculations. With the intercalation of Cu atoms, more electrons fill the bottom of conduction bands of $ZrSe_2$, which change the electronic structure from semiconducting to metallic characters. Notably, this presented intercalation approach does not induce notable changes of crystal structure and overall band dispersions, indicating its high potentials for clean electron doping systems.

A high quality $ZrSe_2$ and Cu intercalated $ZrSe_2$ ($Cu_xZrSe_2$, x = 0.07, x value was determined by ICP) single crystals were synthesized via an improved CVT method. The experimental detail of CVT synthesis is provided in the experimental part of the paper. The single crystal X-ray diffraction (XRD) patterns recorded on $ZrSe_2$ and $Cu_{0.07}ZrSe_2$ samples at room temperature are shown in **Fig. 1**. The (001), (002), (003) and (004) peaks indicate the high quality single crystals of both samples. The diffraction peaks emerged at the same angles in both $ZrSe_2$ and $Cu_{0.07}ZrSe_2$, with a small change in the c-lattice constant from 6.1889Å to 6.2035Å after Cu-intercalation along (001) plane calculated from XRD. Similarly, the peak intensities of $Cu_{0.07}ZrSe_2$ are enhanced, which is due to its increased crystallinity and other growth process. Highly crystalline large-sized single crystals of $ZrSe_2$ and $Cu_{0.07}ZrSe_2$ can be seen inside the XRD figure having a typical lateral size of ~0.8-1cm with silver grey colour and hexagonal shape respectively.

To investigate the chemical compositions and elemental ratio in the grown crystals, X-ray photoelectron spectroscopy (XPS) was performed. Typical XPS spectra of $ZrSe_2$ and $Cu_{0.07}ZrSe_2$ are shown in **Fig. 2**. For $ZrSe_2$, the obvious Zr 3d (Zr-3$d_{5/2}$, Zr-3$d_{3/2}$) peaks at 183.05 eV and 185 eV are displayed, along with clear Se 3d (Se-3$d_{5/2}$, Se-3$d_{3/2}$) peaks at 53.55 eV and 54.2 eV. These peak positions indicate the similar chemical bonding state for both $ZrSe_2$ and $Cu_{0.07}ZrSe_2$, with a small displacement of Zr-3$d_{3/2}$ to lower energy, while donating electron to Se atoms and Se-3d to higher energy due to the emergence of electrons from Zr and Cu atoms in case of $Cu_{0.07}ZrSe_2$. Moreover, **Fig. 2(c)** shows Cu 2p (Cu-2$p_{3/2}$ and Cu-2$p_{1/2}$) peaks at 932.2 eV and 952.2 eV. There is no satellite peak of $Cu^{2+}$ in the obtained $Cu_{0.07}ZrSe_2$ crystals, instead $Cu^{+1}$ peak is visible. This can be ascribed to the electron donation from the intercalated $Cu^o$ atoms to $ZrSe_2$ layers, resulting in $Cu^{+1}$ state that is consistent with previous Cu-substituted compounds.[59]

In order to investigate the detailed local microstructure, compositions and topographical information with an atomic arrangement, we employed a high resolution scanning transmission electron microscopy (STEM) on $Cu_{0.07}ZrSe_2$. **Fig. 3(a)** is typical Z-contrast high resolution image recorded through STEM, showing the surface topography of atoms. It reveals a regular structure with perfect hexagonal atomic arrangement as indicated by inset red line frame. The microstructure of $Cu_{0.07}ZrSe_2$ is similar to that of pristine $ZrSe_2$ (see HRTEM image in **Fig. S1**). **Fig. 3(b)** shows a sketch of elemental configuration of $ZrSe_2$ with a hexagonal arrangements. The layered structure (Zr-Se) is indicating the hexagonal structure of $Cu_{0.07}ZrSe_2$ having two planes of Se atoms separated by Zr atoms with Cu ions randomly intercalated between the layers. Similarly, **Fig. 3(c)** shows a uniform structure of different layers with a d-spacing of about 2.29 Å that corresponds to (002) and Zr-Zr atom bond length of 1.73 Å corresponding to (003) lattice plane. From the d-spacing, it has been observed that $Cu_{0.07}ZrSe_2$ have the same layers spacing as that of $ZrSe_2$, with a negligible enlargement. The Z-contrast intensity of lattices corresponds to the inset side view profile of the chemical structure in **Fig. 3(c)**. The crystal orientation and diffraction pattern were measured by selected-area electron diffraction (SAED). The SAED pattern of $Cu_{0.07}ZrSe_2$ in **Fig. 3(d)** contains a set of diffraction spots along the rotational direction of superlattice (002) and (110) zone axes of a hexagon, which is consistent with XRD results and STEM images. The diffraction spots corresponding to the $Cu_{0.07}ZrSe_2$ are different from pristine sample in terms of the intensities and rational direction (see inset SAED image in **Fig. S1**). The diffraction spots of $Cu_{0.07}ZrSe_2$ are much brighter without $60^o$ rotation in comparison with the lower brightness and have $60^o$ rotation of diffraction spots in $ZrSe_2$. The elemental mapping images in **Fig. 3(e)** further confirm the homogeneous elemental distributions of Zr, Se and Cu atoms in the single crystal $Cu_{0.07}ZrSe_2$. Besides, that the energy dispersive X-ray spectroscopy (EDS) recorded for the samples also shows the existence adequate amount of Cu, Zr and Se elements in the single crystal (see **Fig. S2**, in supporting information). The stoichiometric ratio of Cu atoms is about 7wt%, as determined by inductively coupled plasma mass spectroscopy (ICP-MS) measurements as shown in **Table S1**. Combining the STEM analysis with XPS results, we can suggest that a small amount of Cu atoms have been randomly distributed into $ZrSe_2$ interlayers without any lattice changes.

To study the electronic properties, we performed DFT calculations to explore the band structures of pristine and Cu-intercalated $ZrSe_2$ single crystals. The computational details are given in the experimental section (supporting information). The structural model of bulk $ZrSe_2$ for calculations with a same hexagonal Brillouin zone (BZ) for both $ZrSe_2$ and Cu-intercalated $ZrSe_2$ of high-symmetry points is shown in **Fig. 4(a, b)** respectively. The calculated band structure results for $ZrSe_2$ are depicted in **Fig. 4(c)**. The band structure of $ZrSe_2$ shows that the valance band maximum (VBM) appears at $\Gamma$ point, which is mainly attributed by the p-orbitals of Se atoms. In contrast, the conduction band minima (CBM) appeared at 0.29 eV above the Fermi level at L point, showing a semiconductor with an indirect band gap of 0.71 eV. After the intercalation of Cu atoms, the emergence of conduction band which crosses over the Fermi level at M/L points, indicates the excess electron doping donated from Cu, changing the system from semiconductor to metallic state as shown from the red line in **Fig. 4(c)**, adopted from the calculated band structures of Cu-intercalated $ZrSe_2$ (**Fig. S3(c)**). The donated electron from the Cu 4s shell can make a stable closed-shell $Cu^+$ resulting in an increased charge distribution within $ZrSe_2$ layers. Notably, the intercalation of copper atom merely provides a change of carrier concentration without modifying band dispersions and the structural integrity of $ZrSe_2$ layers. From **Fig. 4(c)** we further confirmed that the emerging CB is composed of $Zr$-$d_{yz}$ and $Zr$-$d_{z2}$ at M and L points, while the VB composed of Se-$p_z$ and Se-$p_y$ of high symmetry Γ point, (as we can see that from **Fig. S3(c)**). The Cu-p and Cu-d can only contribute in deeper energy (**Fig. S3(d)**), which can donate electrons to lift the Fermi level above the bottom of conduction band that is mainly composed of Zr-d orbitals. Based on band calculations, one can notice that a variation of doping can induce either hole or electron-type carriers, as dominated by different orbital characters from distinct elements as shown from **Fig. 4(c)**. This can potentially provide additional freedom to manipulate electronic properties of charge carriers.

We further investigated the band structures through angle resolved photoemission spectroscopy (ARPES). According to previous ARPES investigation of $ZrSe_2$[60], we can determine the $k_z$ value accurately. Using the equation $k_\perp = \sqrt{(2m/\hbar^2)(E_{kin}\cos^2\theta + |V_0|)}$ (where $k_\perp$ is the momentum value along $k_z$, $m$ is the electron mass, $E_{kin}$ is the kinetic energy of the photoelectron,

$V_0$ is the inner potential), in combination with the data in[60], we can determine $V_0 = 10.9$ eV, which gives 30 eV and 42 eV photon energies for $\Gamma$ and $A$, respectively.

**Fig. 5** shows the ARPES intensity of $ZrSe_2$ and $Cu_{0.07}ZrSe_2$ at various binding energies, measured in the first Brillouin zone at T=26 K. There is no spectral weight at the Fermi level in $ZrSe_2$ (**Fig. 5(a₁)**), where one can notice electronic states around M at the Fermi level in $Cu_{0.07}ZrSe_2$ (**Fig. 5(b₁)**), indicating that electrons are doped into the parent electronic system of $ZrSe_2$ after Cu intercalation. As shown in band calculations, these electronic states come from the bottom of conduction bands. While going into deeper binding energies, the spectral weight distributions are very similar in both $ZrSe_2$ (**Fig. 5(a₂₋₅)**) and $Cu_{0.07}ZrSe_2$ (**Fig. 5(b₂₋₅)**), showing a symmetry consistent with the lattice structure of $ZrSe_2$, though the pocket size changes slightly due to different electron doping level.

By comparing the band structure of Cu intercalated $ZrSe_2$ to that of pristine $ZrSe_2$, we can reveal how the electronic structures vary after the intercalation. **Figs. 6(a, b)** show the E vs k dispersions of $ZrSe_2$ and $Cu_{0.07}ZrSe_2$ along the $\Gamma$->M direction, and **Figs. 6(c, d)** show data taken along the A->L direction. The pristine $ZrSe_2$ is a semiconductor, and our ARPES in **Figs. 6(a, c)** show that the Fermi level is within the band gap, and the top of the valence band is located at around -0.87 eV in binding energy. After the intercalation of Cu atoms, the band structures move to deeper binding energy by 70 meV, and the Fermi level cuts across the bottom of conduction bands around the zone boundary (see **Figs. 6(b, d)**), which gives rise to the electron pockets around M shown in **Fig. 5(b₁)**. Such a shift of Fermi level indicates that Cu atoms donate electrons into the pristine $ZrSe_2$ electronic systems, and changes it from a semiconductor to a conductor. There are no evident changes in the band dispersions with the Cu intercalation, suggesting that the electron correlation and the dimensionality of individual bands remain the same as pristine $ZrSe_2$.

In **Figs. 6(e, f)**, one can notice that the bottom of the conduction band is deeper at L, and this difference arises from the 3D characteristics of the conduction bands. The introduction of Cu induces an electron doping without the change of overall band dispersions, and this provides an opportunity to study the gap structure in the pristine ZrSe$_2$. In **Figs. 6(b, d)**, we can clearly see that there is an indirect gap. The top of the valence band is at -0.94 eV, the bottom of conduction band is at -0.17 eV in **Fig. 6(f)**, and these values yield an indirect band gap of 0.77 eV, which is highly consistent with the calculated band gap of 0.71 eV. The red lines in **Figs. 6(e, f)** show the calculated band structures adopted from **Fig. S3(c)** consist of Zr-d$_{yz}$ and Zr-d$_{z2}$ orbitals, which are well matched with the experiment estimation data. In addition, both of **Figs. 6(b, d)** shows non-dispersive spectral weight at around -0.5 eV, which is absent in the pristine ZrSe$_2$. This feature can be attributed to some localized electronic states of Cu atoms intercalated in between the ZrSe$_2$ layers.

Furthermore, electrical properties were also investigated to confirm the metallic character of the as-synthesized Cu$_{0.07}$ZrSe$_2$ in comparison with the semiconducting ZrSe$_2$, the field effect transistors (FETs) were fabricated on ZrSe$_2$ and Cu$_{0.07}$ZrSe$_2$ multilayers (in **Fig. 7a**) using Au/Ti as contact electrodes on SiO$_2$/Si substrate. **Fig. 7(b)** shows the channel length of the FETs with around 10 µm. **Fig. 7(c, d)** represents the output I$_{ds}$-V$_{ds}$ curves of the FET devices based on ZrSe$_2$ and Cu$_{0.07}$ZrSe$_2$. The forward and backward transport behaviour of ZrSe$_2$ displays gate tuneable I$_{ds}$-V$_{ds}$ properties under V$_{gs}$=0V, while the linear curve of the Cu$_{0.07}$ZrSe$_2$ shows metallic behaviour. In **Fig. 7(d)**, the slope of I$_{ds}$-V$_{ds}$ curve of ZrSe$_2$ decreases with decreasing gate voltage from 30 to -30 V, showing n-type transport behaviour (the magnified image from V$_{ds}$ = 0-0.4V as shown in **Fig. S4** with clear gate tuneable characteristics), whereas the linear I$_{ds}$-V$_{ds}$ curves of $Cu_{0.07}ZrSe_2$ for all input gate voltages indicated Ohmic contact like behaviour between the metal electrodes with no tuneable character, strongly confirming the metallic behaviour.

In summary, we have controllably intercalated Cu atoms into layered $ZrSe_2$ single crystal via an improved CVT method, and comparatively studied the changes of microstructure and electronic structure of the $ZrSe_2$ and $Cu_{0.07}ZrSe_2$ by using first-principles calculations and ARPES. Both our DFT calculations and experimental results clearly revealed the presence of the metallic state in $Cu_{0.07}ZrSe_2$ with an indirect bandgap nature that is very different from the semiconducting character in $ZrSe_2$. Similarly, the FET results further confirmed the semiconductor to metallic phase transition in $ZrSe_2$ after the intercalation of Cu atoms. These results are highlighting the significance of atomic intercalation-induced electron doping in 2D layered materials, which might be a vital trend for future valleytronic and electronics applications.

## ASSOCIATED CONTENT

**Supporting Information.** Experimental section, CVT growth procedure, ARPES measurement, theoretical calculations, additional characterization and results.

## AUTHOR INFORMATION

**Corresponding Author**

[*]E-mail: zsun@ustc.edu.cn (Z.S.); song2012@ustc.edu.cn (L.S.)

**Author Contributions**

[+]Z.M. and K.M. contributed equally to this work. L.S. and Z.S. supervised the project and designed the experiments. Z.M. carried out most of the experiments and analyzed the data. K.M. and Z.S. performed ARPES measurement and analysis. H.L. and X.W. performed the DFT calculations. C.W., Z.U.R. and M.H. partially contributed to experimental characterizations. Z.M., K.J.M., Z.S., X.W. and L.S. analyzed the data and co-wrote the paper. All authors discussed the results and commented on the manuscript.

## ACKNOWLEDGMENT

The authors acknowledge the financial support from the MOST (2017YFA0303500, 2017YFA0402901, 2016YFA0200602, 2014CB848900, 2014CB921102), NSFC (U1532112, U1532136, 11574280, 11190022), CAS Key Research Program of Frontier Sciences (QYZDB-SSW-SLH018) and CAS Interdisciplinary Innovation Team. Z.M. acknowledges the CSC (Chinese Scholarship Council) Program. L.S. acknowledges the support from Key Laboratory of Advanced Energy Materials Chemistry (Ministry of Education) Nankai University, and Key Laboratory of the Ministry of Education for Advanced Catalysis Materials and Zhejiang Key Laboratory for Reactive Chemistry on Solid Surfaces (Zhejiang Normal University). We thanks the Hefei Synchrotron Radiation Facility (Angle Resolved Photoemission Spectroscopy and Photoemission Endstations, NSRL), and the USTC Center for Micro and Nanoscale Research and Fabrication for helps in characterizations.

**Notes**

The authors declare no competing financial interests

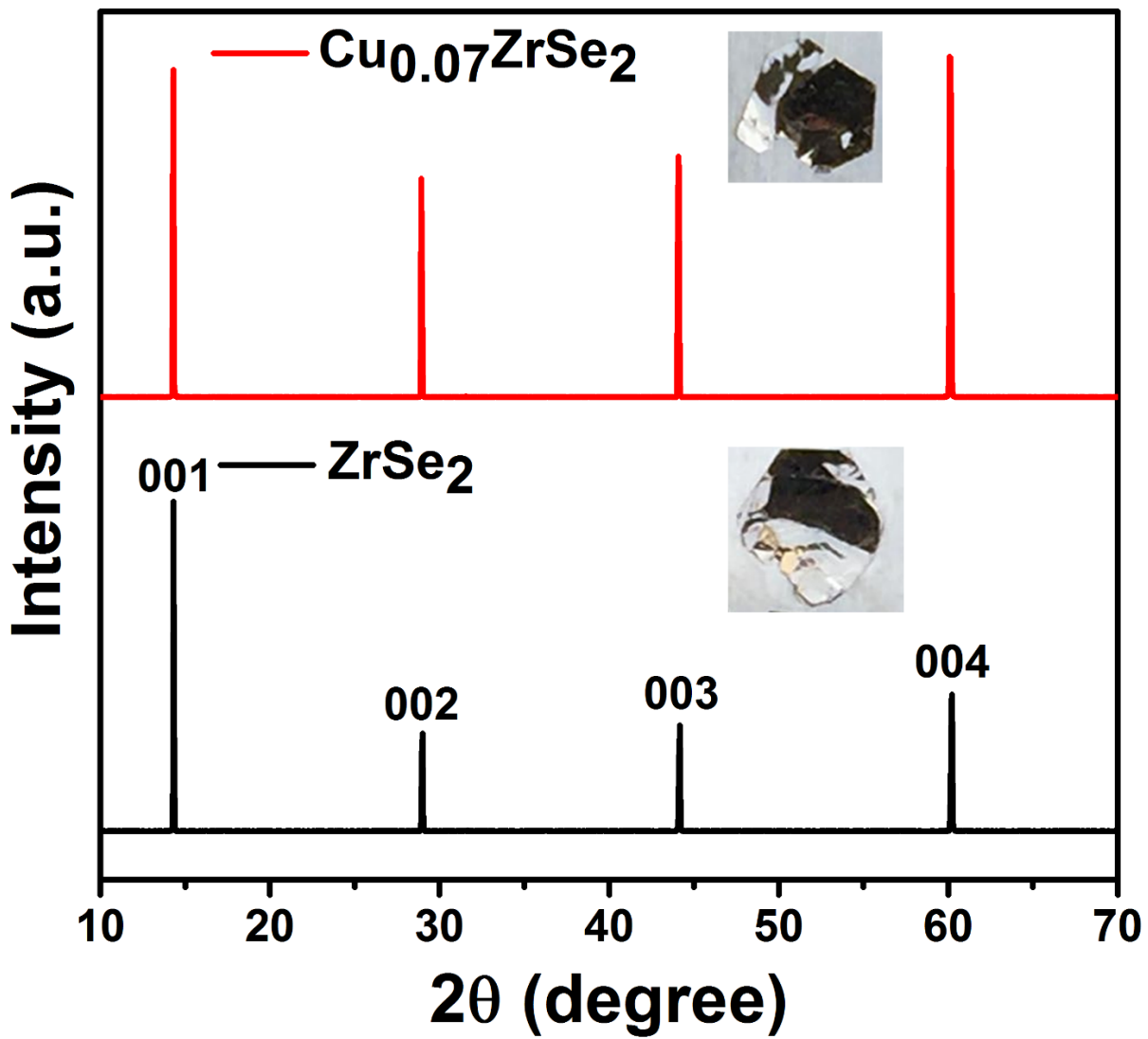

**Figure 1.** Single crystal XRD patterns of ZrSe₂ in comparison with Cu₀.₀₇ZrSe₂ depicting the enhancement of peaks and negligible enlargement.

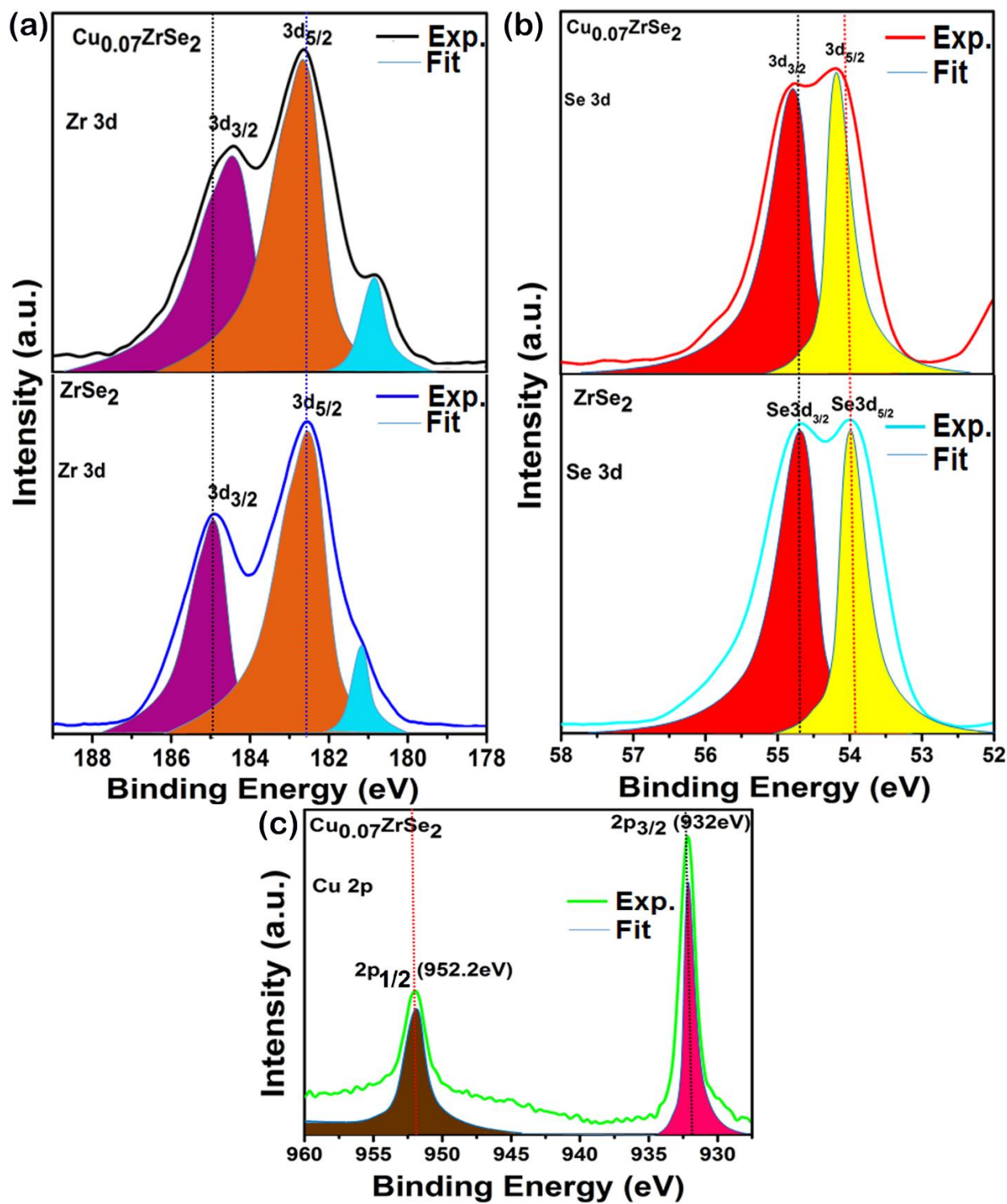

**Figure 2.** XPS analysis of as-synthesized $ZrSe_2$ and $Cu_{0.07}ZrSe_2$ single crystals. (a) Zirconium (Zr 3d). (b) Selenium, (Se 3d). (c) Copper (Cu 2p)

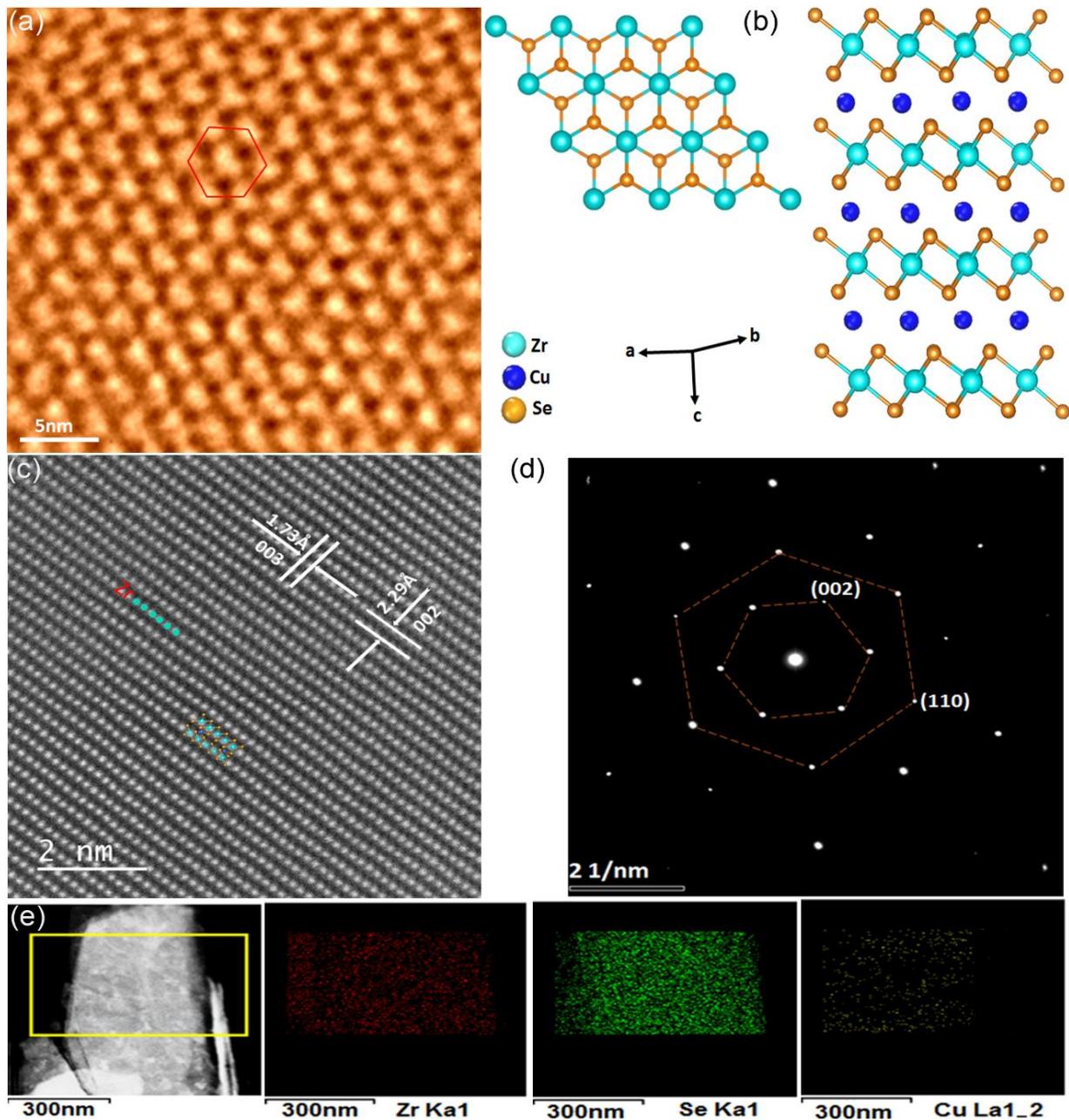

**Figure 3.** Microstructural characterizations of $Cu_{0.07}ZrSe_2$ single crystal through scanning transmission electron microscopy (STEM). (a) Typical STEM image with Z-contrast. (b) Structural configurations of bulk and side view of layered structure with Cu atom between the layers. (c) High resolution STEM image with specified layers identification and (d) SAED patterns of $Cu_{0.07}ZrSe_2$. (e) The elemental mapping image shows the uniform distribution of Zr, Se and Cu elements, respectively.

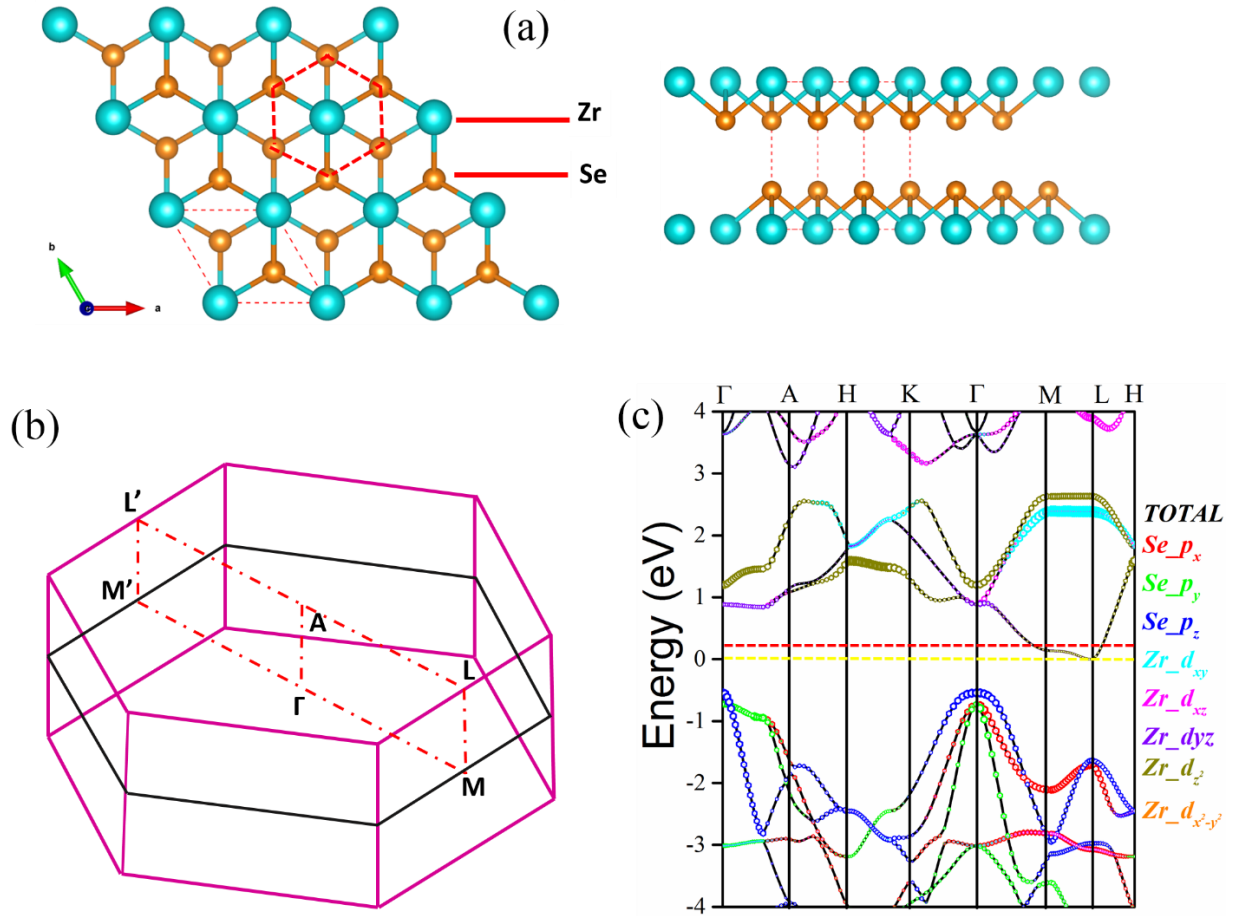

**Figure 4.** (a) Structural model of ZrSe₂. (b) Hexagonal Brillouin zone (BZ). (c) DFT orbital decomposed band structure calculations of bulk ZrSe₂ along the high symmetry k-points. After the intercalation of Cu atoms, the emergence of conduction band at M points indicates excess electrons donation that can change the system from semiconductor to metallic state as shown from the dash red line.

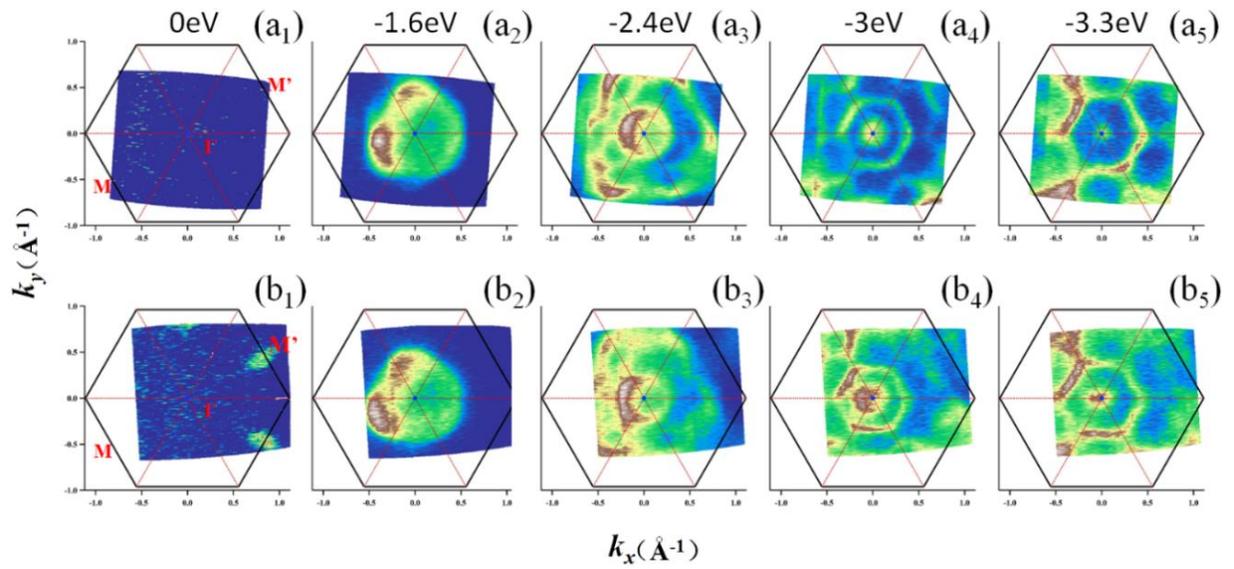

**Figure 5.** ARPES intensity of the pristine $ZrSe_2$ (upper panels, $a_1$-$a_5$) and $Cu_{0.07}ZrSe_2$ (lower panels, $b_1$-$b_5$) single crystals at various binding energies, taken with 30 eV photons. The black hexagonal boxes indicate the Brillouin zone.

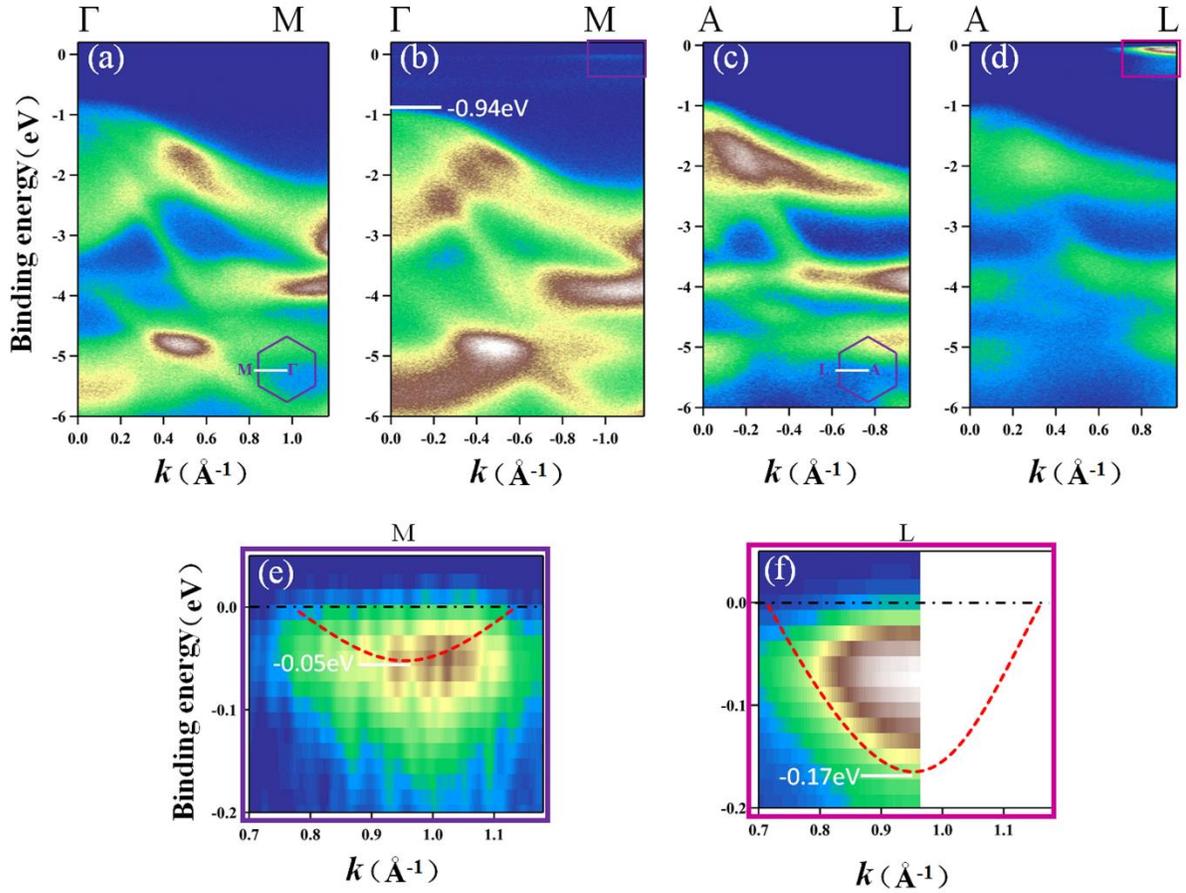

**Figure 6.** (a, b) The photoemission intensity plots along Γ->M direction of pristine ZrSe₂ and Cu intercalated ZrSe₂ single crystals, respectively. The data were taken using 30 eV photons. (c, d) The photoemission intensity plots along A->L high direction of pristine ZrSe₂ and $Cu_{0.07}ZrSe_2$, respectively, taken using 42 eV photons. (e, f) The spectral weight of conduction bands around the zone boundary in panels b and d, respectively. The red dashed lines indicate the band dispersions from DFT calculations (adopted from Figure 4 (c)), consists of Zr $d_{yz}$ and $d_{z2}$ orbitals.

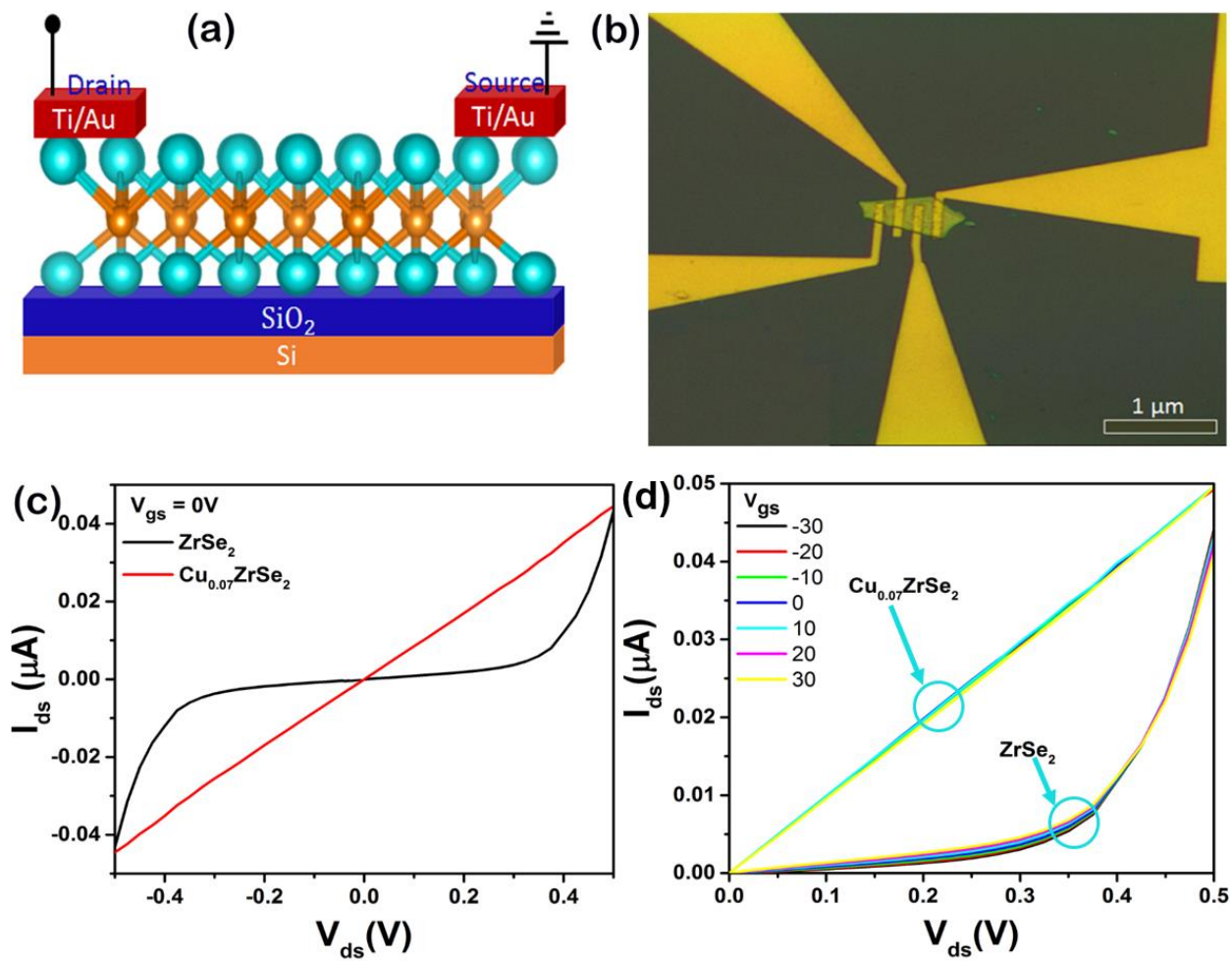

**Figure 7.** Typical FET device fabricated on ZrSe$_2$ and Cu$_{0.07}$ZrSe$_2$ exfoliated flakes. (a) Schematic of the device structure. (b) Optical image of the as-shaped nanodevice. The current-voltage curves at gate voltage = 0V showing (c) the forward and backward transport behavior indicate semiconducting and metallic behavior for ZrSe$_2$ and Cu$_{0.07}$ZrSe$_2$, respectively. (d) The output characteristics curves of ZrSe$_2$ and Cu$_{0.07}$ZrSe$_2$ at different gate voltages.

**The table of contents:**

**Single crystal ZrSe₂ and atomic Cu-intercalated ZrSe₂ layered structures** were synthesized in high crystalline quality by using chemical vapor transport method. Both theoretical DFT calculations and experimental ARPES results revealed a tramsition from semiconducting to metallic characteriztics, indicating the emergence of conduction band dispersion at M/L point of Brillouin zone due to the electron doping of ZrSe₂ layers originated from additional atomic Cu intercalation.

*TOC Figure:*

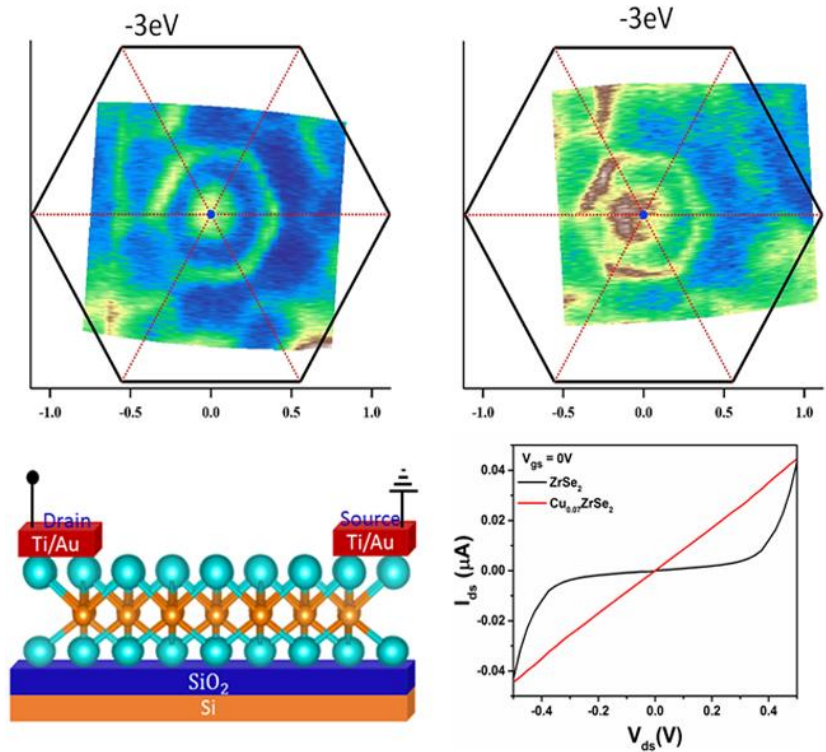